\documentclass[fleqn,usenatbib]{mnras}
\usepackage{newtxtext,newtxmath}
\usepackage[T1]{fontenc}
\usepackage{ae,aecompl}
\usepackage{rotating}
\usepackage{lscape}
\usepackage{float}
\usepackage[section]{placeins}
\usepackage{rotating}
\usepackage{lipsum}

\usepackage{graphicx}	
\usepackage{amsmath}	
\usepackage[mathscr]{euscript}
\usepackage{tablefootnote}
\usepackage{subfig}

\title{A Search for Pulsar Companions Around Low--Mass White Dwarfs}
\author[T. M. Athanasiadis et al.]{
Tilemachos M. Athanasiadis,$^{1}$\thanks{E-mail:tilemathan@mpifr-bonn.mpg.de}
Marina Berezina,$^{1,2}$ 
John Antoniadis$^{1,3,4}$
\newauthor David J. Champion,$^{1}$ 
Marilyn Cruces,$^{1}$ 
Laura Spitler,$^{1}$
and Michael Kramer$^{1}$
\\
$^{1}$Max-Planck-Institut f\"{u}r Radioastronomie (MPIfR), Auf dem H\"{u}gel 69, 53121, Bonn, Germany\\
$^{2}$Landessternwarte, Universit\"{a}t Heidelberg, K\"{o}nigstuhl 12, 69117, Heidelberg, Germany\\
$^{3}$Argelander Institut f\"{u}r Astronomie, Auf dem H\"{u}gel 71, 53121, Bonn, Germany\\
$^{4}$Institute of Astrophysics, FORTH, Dept. of Physics, University of Crete, Voutes, University Campus, GR-71003 Heraklion, Greece
\\
}

\date{Accepted 2021 May 27. Received 2021 May 27; in original form 2021 February 11}

\pubyear{2021}

\begin{document}
\label{firstpage}
\pagerange{\pageref{firstpage}--\pageref{lastpage}}
\maketitle

\begin{abstract}
We report on a search for pulsars at the positions of eight low--mass white dwarfs and one higher--mass white dwarf with the 100--m Effelsberg Radio Telescope.
These systems have orbital parameters suggesting that their unseen companions are either massive white dwarfs or neutron stars. 
Our observations were performed at 1.36\,GHz, reaching sensitivities of 0.1--0.2\,mJy. 
We searched our data accounting for the possible acceleration and 
jerk of the pulsar signals due to orbital motion, but found no 
significant pulsar signals.
Considering our result jointly with 20 non--detections of similar systems with the Greenbank Radio Telescope, we infer $f_{\rm NS}\leq 0.10$, for the fraction of NSs orbiting these white dwarfs.
We discuss the sensitivity of this result to the underlying assumptions and conclude with a brief discussion on the prospects of targeted surveys for discovering millisecond pulsars. 
\end{abstract}


\section{Introduction}
\label{sec:intro}
The formation of white dwarfs with low masses ($<0.3$\,M$_{\odot}$; henceforth low--mass white dwarfs; LMWDs) is thought to require mass transfer in a multiple system, as most low--mass isolated stars in our Galaxy have not yet evolved off the main sequence \citep{marsh1995}. 
Indeed, spectroscopic studies \citep[e.g.][and references therein]{2020brown} suggest that most LMWDs have dark, compact companions. 
The majority of these stars are thought to be carbon-oxygen WDs \citep{2007leeuwen,2009aguerosa,Andrews2014,Boffin2015}, however neutron star (NS) and black hole (BH) companions are also possible.
The primary channel for the formation of the latter, is long-term mass accretion from the LMWD progenitor onto the first-born compact object. For NS primaries, the spin-up due to gain of angular momentum can lead to ``recycling'', i.e. the formation of a rapidly-spinning millisecond pulsar \citep[MSP;][]{1991Bhattacharya}. 

While compact MSP/LMWD binaries are extremely valuable for a broad range of physical inquiry,  only a handful are known so far \citep{Antoniadis:2014eia,Wex:2020ald}. 
The advent of systematic spectroscopic WD  surveys  \citep[][and references therein]{Eisenstein:2006ty,brown2010,Kilic:2010ws,2020brown} provides a unique opportunity to
search for MSPs among the companions of nearby LMWDs. Pulsars discovered in this way can contribute precise NS mass measurements, which in turn are important for 
constraining the equation-of-state of nuclear matter \citep[][]{Antoniadis2013,Cromartie:2019kug} and strong-field gravity \citep{Wex:2020ald}. 
In addition, understanding the population of LMWD binaries is important for probing various aspects of stellar evolution \citep[e.g.][]{2014istrate}, as well as the formation and properties of double-degenerate binaries in the Galaxy \citep[e.g.][]{2019Li}. 

One method that has been employed in the past for the detection of binary MSPs,  is the targeted search for pulsars at the positions of known LMWDS. Several studies, \citep[e.g. see]{2007leeuwen,2009aguerosa,2020brown}, have followed-up tens of LMWDs with deep radio observations in search for pulsar signals. These surveys have thus far yielded null results, setting an upper limit for the fraction of LMWDS orbited by NSs to $\lesssim 10-15\%$ \citep{2009aguerosa}.
In this paper, we expand on these previous surveys by reporting the results of a targeted radio survey for MSPs  with the 100-m Effelsberg radio telescope, at the positions of eight binary LMWDs and one higher-mass WD \citep{brown2013,2020brown}. 
Based on their mass functions, all these systems have massive  ($\gtrsim 0.8$\,M$_{\odot}$) companions that could potentially be MSPs. 
The paper is organized as follows: In Section~\ref{sec:systems_observations}, we present our target selection and observations.
Section~\ref{sec:results} describes our data analysis and results. We conclude with a discussion in Section~\ref{sec:discussion}.

\section{Observations and data reduction}
\label{sec:systems_observations}
\subsection{Target selection}
\label{sec:target_selection}
Our targets are binary WDs with dark companions, discovered and characterized by the Extremely Low Mass White Dwarf Survey   \citep[henceforth ELM survey][]{brown2010,2020brown}, a systematic spectroscopic survey of nearby ($\lesssim 2$\,kpc) LMWDs. 
Targets are selected based on their photometric colors \citep{brown2010}. Spectra are then obtained at a few epochs, looking for radial velocity changes due to orbital motion. 
After being identified, binaries are followed up  until the orbit is characterized to sufficient precision.
Therefore, for each LMWD binary, the survey provides the mass function, $ f = {P_{\rm orb}\,{K^3}}/{2\pi G}$, 
where $P_{\rm orb}$ is the orbital period, $K$, the semi-amplitude of the WD orbital radial velocity, and $m_{\rm WD}$, an estimate of the WD mass obtained by modeling the Balmer lines of its optical spectrum.
Combined with the orbit inclination, $i$, the former estimates yield the mass of the companion star, $m_{\rm c}$, using Kepler's laws, $M_{\rm tot}^2 = m_{\rm c}\sin i / f$, where  $M_{\rm tot}=m_{\rm c}+m_{\rm WD}$.
While the inclination is unknown, assuming an edge-on orbit ($i=90^{\rm o}$) yields 
a minimum value for the companion mass. Similarly, adopting a probability distribution for the inclination that is uniform in $\cos i$, one can obtain a probability distribution function (PDF) for $m_{\rm c}$.

\begin{figure}
    \centering
    \includegraphics[width=\columnwidth]{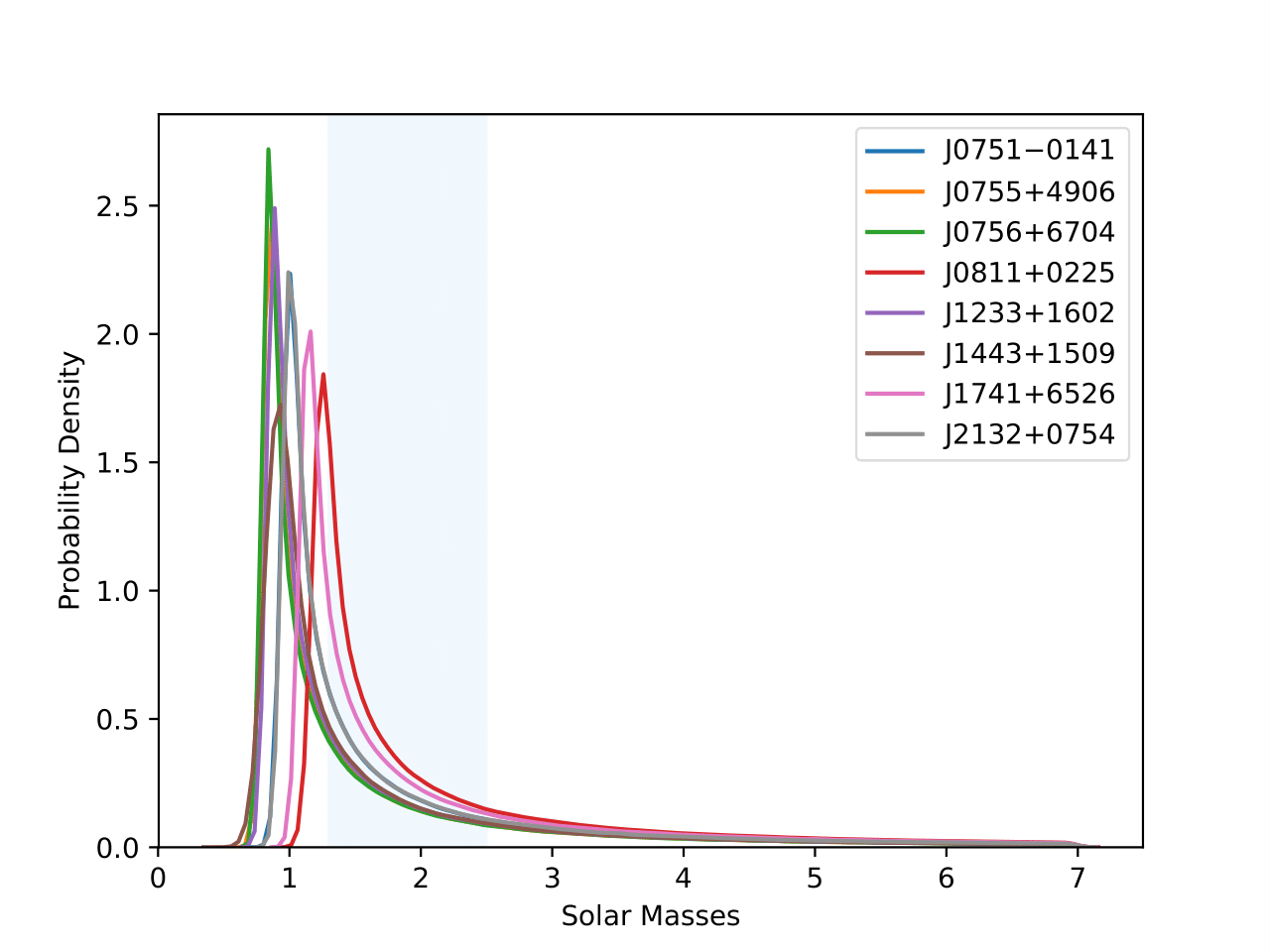}
    \caption{Probability Density Function (PDF) for the companion mass of the 8 targets observed with Effelsberg. The shadowed area shows the mass range for which a companion is assumed to be a NS. }
    \label{fig:pdfs}
\end{figure}

For our observations, we selected WDs with minimum companion masses greater than 0.8\,M$_{\odot}$. The companion-mass PDFs for these systems are shown in Figure~\ref{fig:pdfs} and 
their properties are  presented in Tables~\ref{tab:acc_jerk_dm_tobs} and \ref{tab:object_priors}. As can be seen in Table~\ref{tab:object_priors}, 
all our targets have compact orbits ($P_{\rm orb} \leq 20$\,h) and a relatively high 
probability of hosting a NS companion, $P_{\rm prior}^{\rm NS}\gtrsim 0.21$, based on 
the m$_{\rm c}$ PDFs. Here, for simplicity, to infer the prior probability of a given system to host a NS, from the  m$_{\rm c}$ PDF,  we assume a sharp dividing line between 
WDs and NSs (at 1.3\,M$_{\odot}$), and NSs/BHs (at 2.5\,M$_{\odot}$). This choice does not affect our ability to detect pulsars, but does influence somewhat our analysis presented in Section~\ref{sec:results}. A more detailed discussion of this assumption as well as alternatives follows in Section~\ref{sec:discussion}. 

Optical observations also place 
constraints on a number of quantities 
that are useful for radio follow-up,  
including the distance,  
dispersion measure (DM), and the 
amplitudes of the radial orbital 
acceleration and its first derivative \citep[jerk, see][]{2004handbook}:
\begin{equation}
\label{eq:acc}
    a_{\rm max}=\pm(G{\rm M}_{\odot})^3\,\frac{m_{\rm WD}}{(m_{\rm WD}+m_{\rm c}^{\rm min})^{\frac{2}{3}}}\,\Omega_{\rm orb}^{4/3}
\end{equation}
and
\begin{equation}
\label{eq:jerk}
    j_{\rm max}=\pm(G{\rm M}_{\odot})^3\,\frac{m_{\rm WD}}{(m_{\rm WD}+m_{\rm c}^{\rm min})^{\frac{2}{3}}}\,\Omega_{\rm orb}^{7/3}
\end{equation}
where $G$ is the gravitational constant and $\Omega_{\rm orb}$ is the orbital angular frequency. 

Distances to most targets (see Table 
\ref{tab:object_priors})  were inferred using 
the \emph{Gaia}\,Early Data Release 3\,(EDR3) parallax estimates \citep{Lindegren:2018cgr,bailer-jones2018}. 
Full posterior distance likelihoods were 
obtained following the method discussed in 
\cite{Antoniadis:2020gos} and references 
therein.  A comparison of our $2-
sigma$ error bounds with the distance 
estimates provided by \citet{2021bailer-jones,vo:eDR3_lite_dist} 
showed that for our targets, the two parallax 
inversion methods yield similar results that 
are consistent within  $\sim10\%$.
For targets lacking reliable astrometric 
solutions, we used photometric distance 
estimates provided by \citet{vo:eDR3_lite_dist} 
catalog. For the distance likelihood distribution, we  assumed a normal distribution with a dispersion equal to the  
$68\%$ confidence interval from the 
same catalogue.

Distance estimates combined with a model for the free electron density in the Galaxy, also yield the expected DM range. For the values given in Table\,\ref{tab:object_priors}, we employed the NE2001 electron density model assuming a 20\% systematic uncertainty \citep{NE2001}. 

\begin{table}
\begin{center}
\begin{tabular}{llllll}
\hline
\hline
Object      & accel. & jerk     & DM range & Tobs \\
            &(${\rm m}/{\rm s}^2$) & (${\rm cm}/{\rm s}^3$) & (${\rm pc}\,{\rm cm}^{-3})$ & (hours) \\
\hline
J0751--0141  & 71.47  & 6.5  & 34.2 -- 54.4  & 1.5  \\
J0755+4800  & 11.89  & 0.16  &  1.8 --  1.9  & 2.5  \\
J0755+4906  & 105.41 & 12.16 & 11.6 -- 55.0  & 1.5  \\
J0756+6704  & 5.09   & 0.06  & 40.7 -- 44.7  & 1.0  \\
J0811+0225  & 2.75   & 0.02  & 25.5 -- 51.7  & 2.0  \\
J1233+1602  & 32.05  & 1.54  &  7.7 --  9.7  & 2.0  \\
J1443+1509  & 23.95  & 0.91  &  4.7 --  9.3  & 2.6  \\
J1741+6526  & 92.05  & 10.95 & 16.7 -- 28.0  & 1.5  \\
J2132+0754  & 15.41  & 0.45  & 17.9 -- 29.1  & 3.0  \\
\hline
\end{tabular}
\caption{Maximum acceleration and acceleration derivative (jerk) ranges for our targets,  inferred using equations \ref{eq:acc} and \ref{eq:jerk}.
The expected DM ranges were calculated using the the distances presented in Table \ref{tab:object_priors} based on the \texttt{NE2001} model \citep{NE2001}. \label{tab:acc_jerk_dm_tobs}}
\end{center}
\end{table}

\subsection{Observations}
\label{sec:observations}
Our targets were observed with the 100-m Effelsberg radio telescope and its dual-polarization 7-beam receiver \citep{2013barr} during several sessions in May 2014 and April 2020. We recorded data over a bandwidth of 300\,MHz (effective bandwidth of 240\,MHz), with the receiver centered at 1360\,MHz. 
The data were sub-divided into 512$\times 0.58$\,kHz channels with a digital filterbank, and finally integrated every 54.6113\,$\mu$s with a Pulsar Fast Fourier Transform Spectrometer (PFFTS) backend. They were stored as 32-bits and then down-sampled into 8-bit format.

Adopting a modified radiometer equation suitable for pulsed signals \citep{2004handbook}, and assuming a mean system temperature of T$_{\rm sys}=29$\,K, a Gain of $1.5$\,K\,Jy$^{-1}$ \citep{berezina2019}, and a typical integration time of $t_{\rm obs}=0.5$\,h, the minimum detectable flux corresponding to S/N = 7 for this setup is $\sim0.1$\,mJy for a  signal with a spin period of 5\,ms, a duty cycle of 33\% and a DM of 20\,pc\,cm$^{-3}$. However, as we discuss below, the true sensitivity appears to be affected due to the presence of radio frequency interference (RFI).

\begin{center}
\begin{figure}
	\includegraphics[width=\columnwidth]{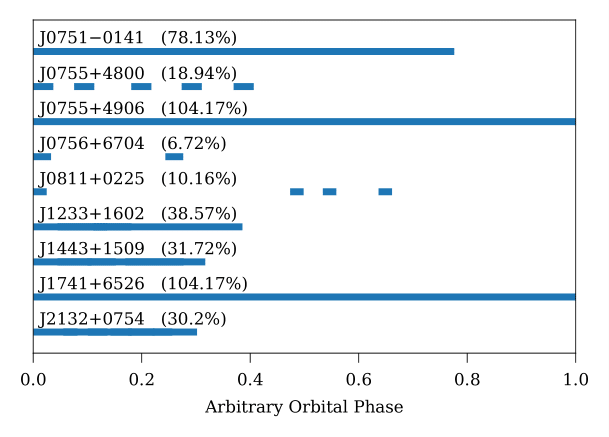} 
    \caption{Orbital phase coverage for each target, calculated using the available ELM survey ephemerides. $\phi=0$ corresponds to the time of ascending node for the WD, $T_{0}^{\rm WD}$. However, we note that due to large uncertainties in the reported  $T_{0}^{\rm WD}$ values, the true orbital phases at the time of our observations may differ substantially. For  J0756+6704,  $T_{0}^{\rm WD}$ was not available and therefore $\phi=0$ corresponds to the start of our first observation.}  
    \label{fig:phasecoverage}
\end{figure}
\end{center}

The three most compact binaries (J0751$-$0141, J0755+4906 and J1741+6526)  were observed  only once, for 1.5\,h each, while the remaining targets were monitored multiple times in 0.5-h sessions to cover large parts of their orbits (see Figure\,\ref{fig:phasecoverage} for the orbital coverage per target).  Observations of the bright pulsar B0355+54 obtained at the beginning of each run were used as a crosscheck that the system was working properly. 

\subsection{Data analysis}
\label{data_search}
Analysis was performed using MPIfR's \textsc{hercules} cluster at the Max Planck Computing and Data Facility in Garching. 

The first stage in our processing  was the filtering of RFI. For this, we used the \texttt{RFIFIND} routine of \texttt{PRESTO} \citep{Ransom2001}, 
as well as  a multi-beam RFI mitigation technique \citep{cruces2020}.
The latter was based on the fact that  RFI is usually detected by multiple beams.  
Hence, if a candidate appeared in at least four of the seven beams of the receiver, it was flagged as RFI. 
Following RFI mitigation, we mainly focused on  data taken with the central beam which contained the sources of interest, but the other  beams were also searched for  serendipitous discoveries. 

For the rest of the analysis we used a hybrid \texttt{SIGPROC}- \citep{sigproc2011} and \texttt{PRESTO}- based pipeline \citep{Ransom2001}. The central-beam data were de-dispersed at 610 
trial DM values, between 0 and 500 pc\,cm$^{-3}$ (thereby cover the expected DM range through the entire  galactic plane). The optimal number of trials was calculated using the \texttt{DDplan} tool of \texttt{PRESTO}. 
The de-dispersed time series were then searched for periodic signals using fast Fourier transforms (FFT) and incoherent harmonic summing. 
To account for the possible smearing of the Fourier power due to orbital motion, we resampled the time series assuming that the source of the signal was subjected to a constant acceleration --- a method known as acceleration search. 
The number of trial acceleration values depends on the acceleration range (Table \ref{tab:acc_jerk_dm_tobs}) and was calculated as in \citet{Eatough2013}.
For circular binaries such as our targets, the assumption of constant acceleration is valid if the observing time does not exceed $\sim$10\% of the orbital period. 
For the cases that the observation time was longer, the first acceleration derivative (jerk) was also taken into account with trial values also based on \citet{Eatough2013}. 
Overall, the former procedures enabled us to coherently search segments of up to $\sim$20\% of the orbital period, i.e. 0.4, 0.3 and 0.3\,h for J0751--0141, J0755+4906 and J1741+6526 respectively.   

Candidate periodic signals with a signal-to-noise ratio higher than 7 were vetted  by eye, but no pulsar was identified. 
We also applied a complementary ``segmented'' search \citep{NG2015}. Following this method, the observations were split into shorter, equal segments, each covering up-to 10\% of the corresponding orbital period of the system. Each part  was then searched, accounting only for constant acceleration. This was done for two reasons: 
firstly, this method makes fewer assumptions about the orbit. 
secondly, acceleration-jerk 
searches return many candidates 
due to the high number of trials 
and therefore, a real signal could be missed due to artifacts. No 
candidate was detected during this stage either. Finally, we did not detect any pulsars in the rest of the beams.

To summarise, we report no detection and are confident, given the multiple observations of each target, that no pulsar was missed due to orbital smearing or scintillation. Due to the nature of these systems, we do not expect to have missed any signals due to radio eclipses either. 
In the next Section, we use these non-detections to place constraints on the fraction of LMWDs orbiting NSs (f$_{\rm NS}$).

\subsection{Survey Sensitivity}
\label{sec:true_sensitivity}
In order to understand the astrophysical implications of our results (see Section \ref{sec:results}) we need to investigate the ability of our survey to detect pulsars.
This ability is directly correlated with the sensitivity of the survey, which we examine here in detail.

As mentioned in Section\,\ref{sec:observations}, the minimum detectable flux for our survey predicted by the radiometer equation is $\sim0.1$\,mJy. However, one expects sensitivity losses due to the effective shortening of the bandwidth and  observing length from  RFI. Some Fourier frequency bins may also affected by  residual periodic RFI that was not filtered out in during the data  reduction. 
 To obtain a more robust estimate of the sensitivity as a function of spin period and DM, we followed the method outlined in \citet{Lazarus2015}. More specifically, we 
injected fake pulsar signals in our data with spin periods ranging from 1 to 30\,ms 
and target-specific DM values within the ranges given in Table\,\ref{tab:acc_jerk_dm_tobs}. The  
routine used in this analysis \citep{Lazarus2015} requires the expected SNR (or equivalently flux density) as an input. Here, we relied on the results of the   detailed sensitivity analysis of the 7-beam receiver described in \citet{berezina2019}. This work, based on re-detections of 165 known pulsars concluded that the sensitivity of the central beam in the absence of RFI is very close to the one predicted by the radiometer equation (Section~\ref{sec:observations}).

Following calibration of the injected signal, we run the data through our pipeline and 
recovered the signal with SNR that was  different from the  one injected. The true 
sensitivity at a given period and DM was obtained by varying the flux of the injected signal until it fell below the detection threshold (at SNR$_{\rm true}\leq 7$). 
This analysis revealed that sensitivity generally varied between different sessions, even for the same object.  
For specific spin periods, the true sensitivity  appears to be a factor of $\sim 1.4$ worse than the theoretical estimate.

\section{Results}
\label{sec:results}
\par The prior knowledge of each system's properties provided by the ELM survey combined with the robust characterization of our survey's sensitivity described above, allows us to place constraints on the  number of NSs in our sample and consequently the  fraction of Galactic LMWDs,  $f_{\rm NS}$,  orbited by NSs.
This quantity depends on: 

\begin{enumerate}
    \item The prior probability of a given system to host a NS, P$_{\rm prior}^{\rm NS}$, based on the mass PDF of the unseen companion (Figure\,\ref{fig:pdfs}). 
    As discussed above, a simple, conservative estimate for  P$_{\rm prior}^{\rm NS}$ can be obtained by assuming that all objects with masses between 1.3 and 2.5\,M$_{\odot}$ are NSs (see  Table\,\ref{tab:object_priors}), 
    
    \item the sensitivity of the survey (see Section\,\ref{sec:true_sensitivity}), and 
    
    \item the probability of a NS in a given system to appear as an MSP, which in turn depends on the MSP luminosity distribution and beaming fraction \citep{2004handbook}.
\end{enumerate}

\begin{figure}
\centering
\begin{tabular}{ccc}
\subfloat[J0751--0141]{\includegraphics[width = 0.4\columnwidth]{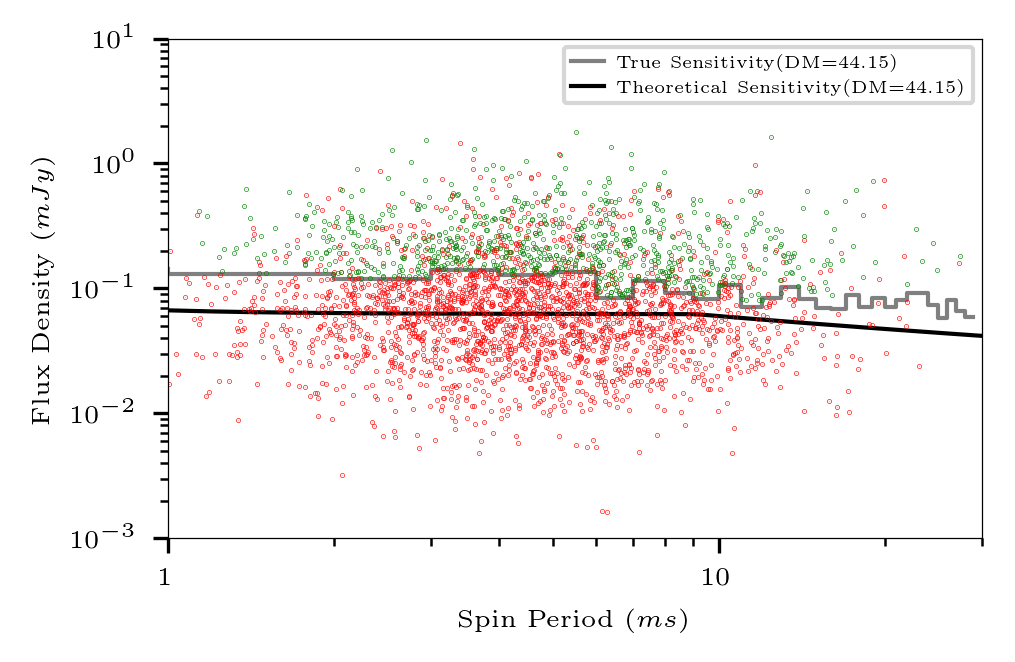}} &
\subfloat[J0755+4906]{\includegraphics[width = 0.4\columnwidth]{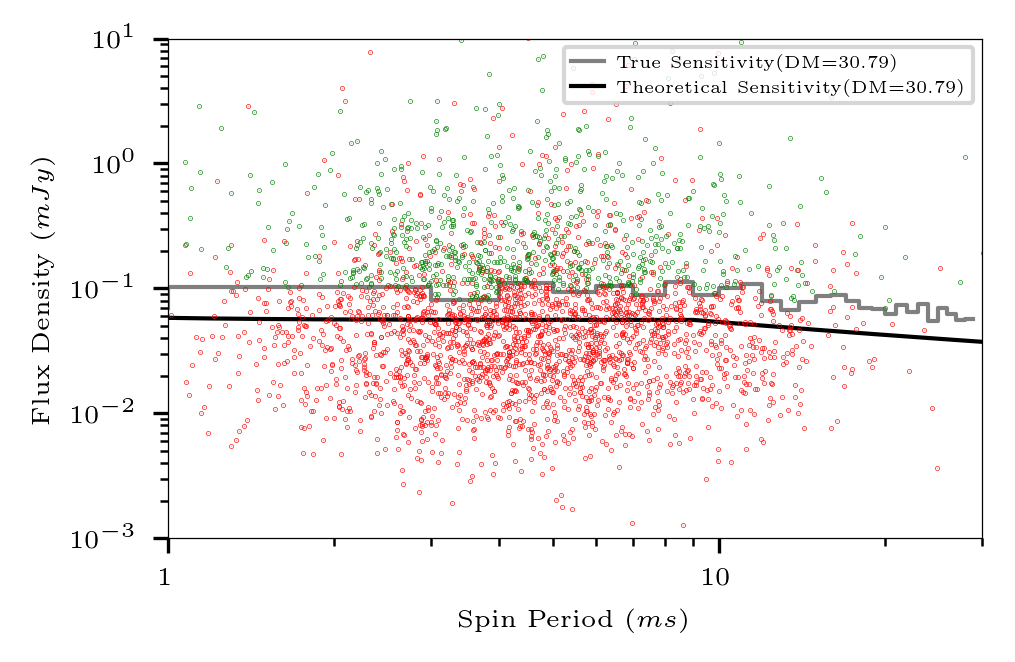}}\\
\subfloat[J0756+6704]{\includegraphics[width = 0.4\columnwidth]{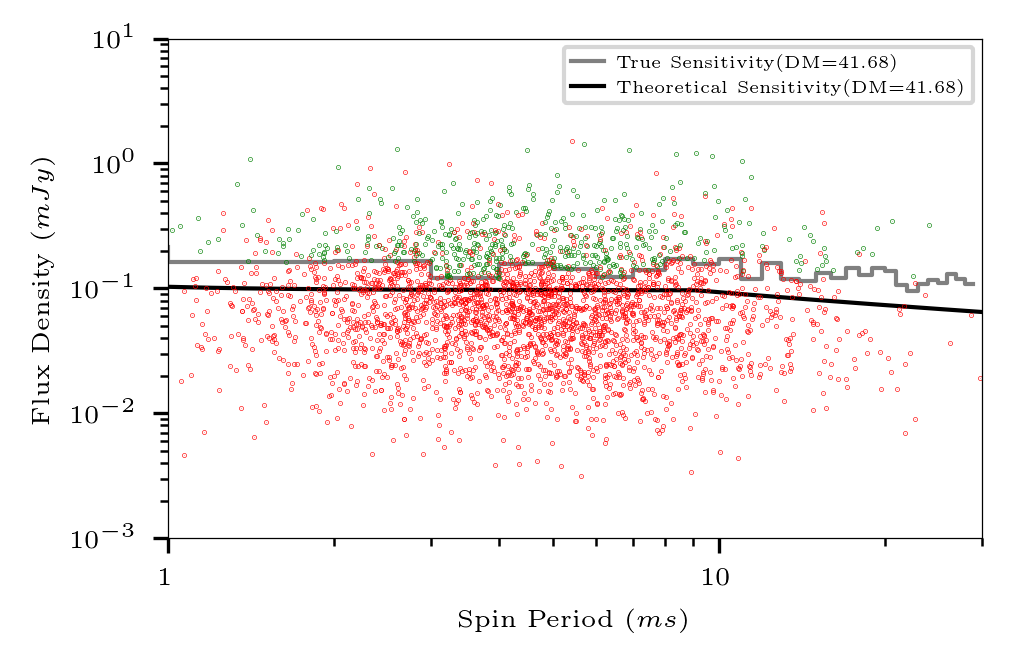}} &
\subfloat[J0811+0225]{\includegraphics[width = 0.4\columnwidth]{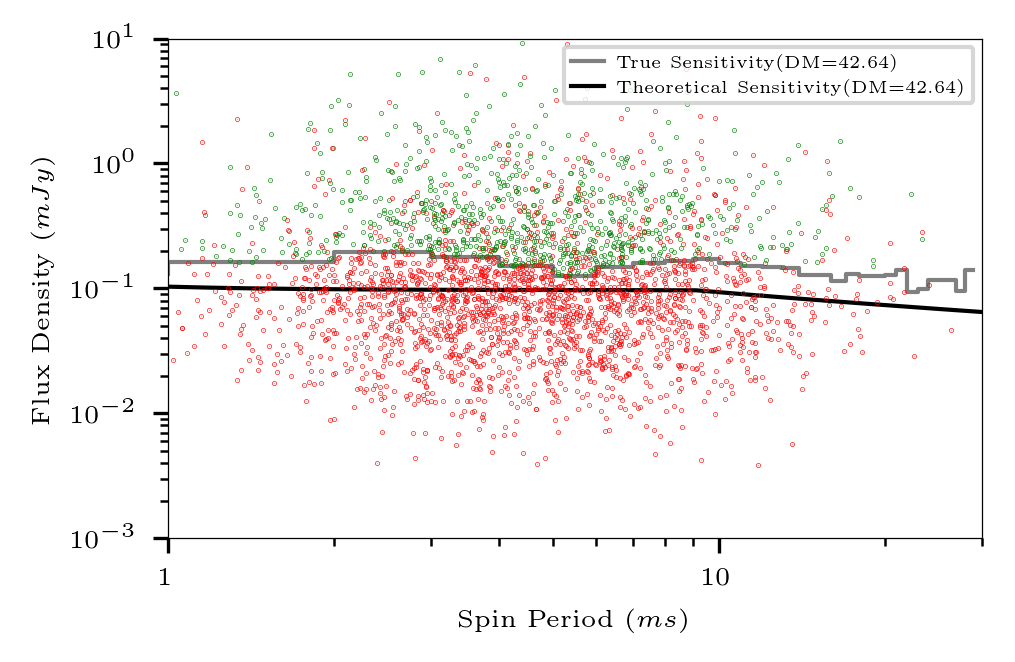}}\\
\subfloat[J1233+1602]{\includegraphics[width = 0.4\columnwidth]{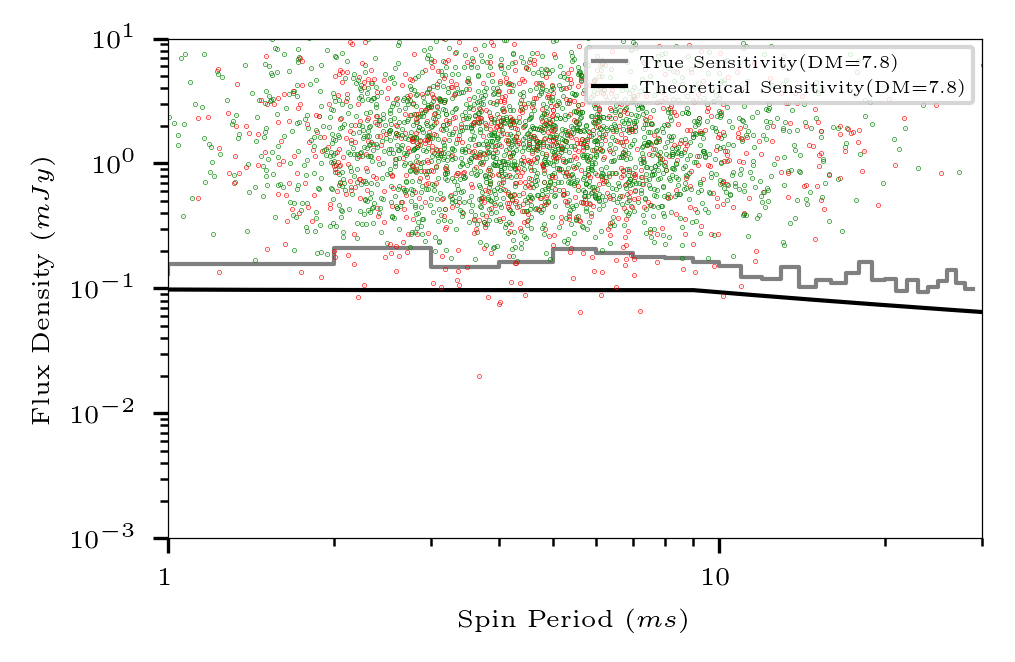}} &
\subfloat[J1443+1509]{\includegraphics[width = 0.4\columnwidth]{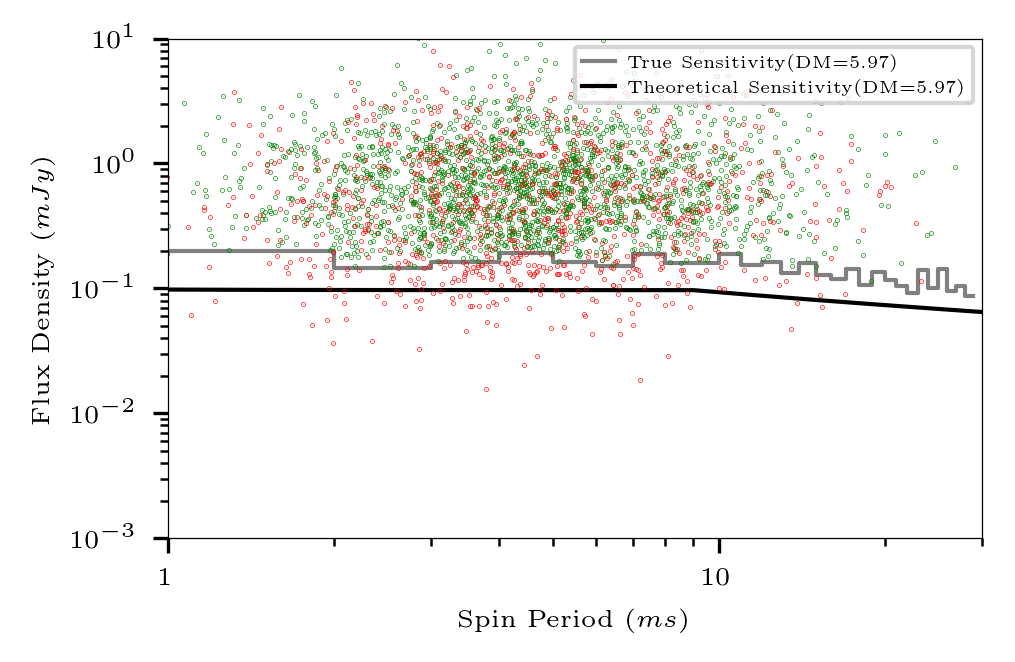}}\\
\subfloat[J1741+6526]{\includegraphics[width = 0.4\columnwidth]{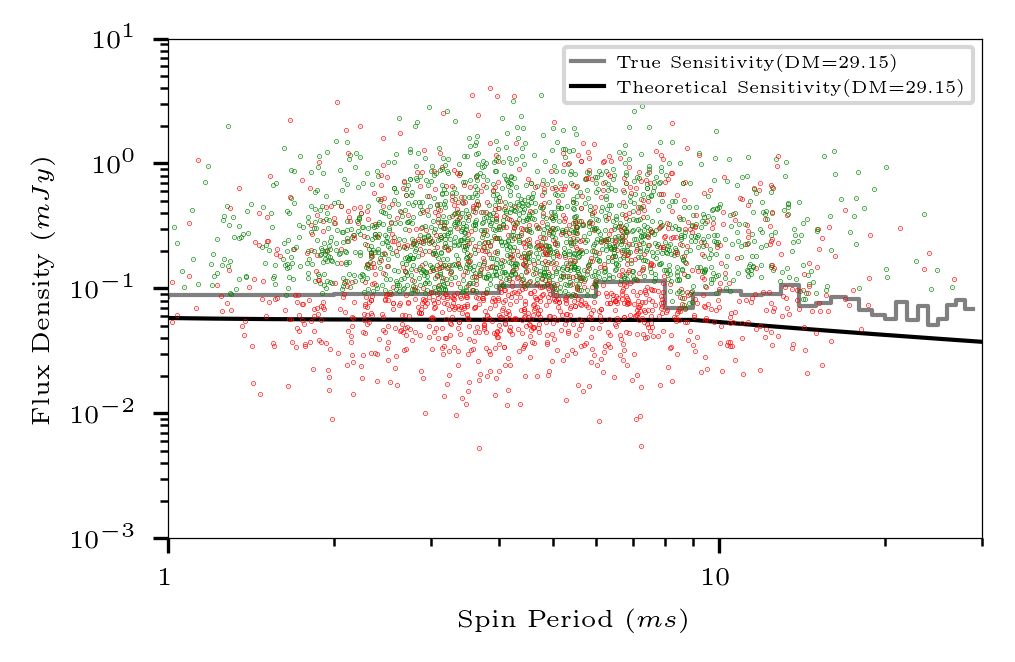}} &
\subfloat[J2132+0754]{\includegraphics[width = 0.4\columnwidth]{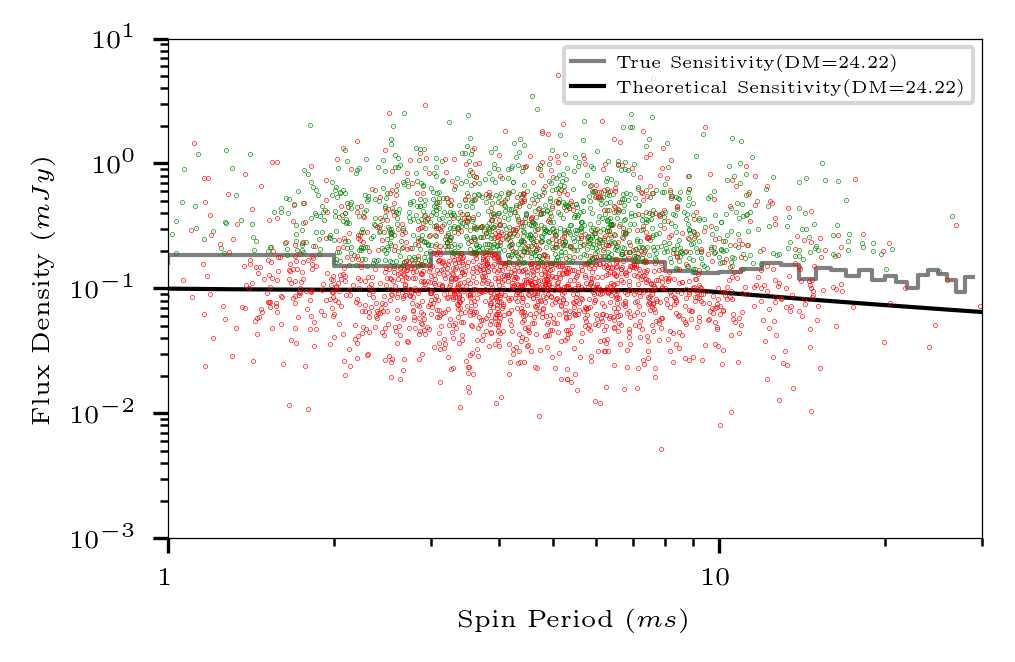}}\\
\end{tabular}
\caption{Simulated MSPs that would have been detected or not (green and red circles respectively) due to sensitivity (below the sensitivity limit) or due to beaming (red dots above sensitivity line). The solid line corresponds to the theoretical sensitivity at the specific DM while the dashed line shows the actual sensitivity calculated as we describe in Section\,\ref{sec:true_sensitivity}}
\label{fig:det_plots}
\end{figure}

To evaluate these factors and place constraints on  $f_{\rm NS}$, we employed a simple Monte-Carlo (MC) sampling scheme that works as follows: 
In each MC realization we begun by sampling the companion mass PDF of each target. If the companion mass is consistent with a NS, we then assigned a spin period and radio luminosity drawn from log-normal distributions \citep{Lorimer2015,2006FaucherKaspi} given by:

\begin{equation}
    {\rm PDF}(x, \mu, \sigma^2) = \frac{1}{x\sigma\sqrt{2\pi}}\exp\left( -\frac{(\log(x)-\mu)^2}{2\sigma^2}  \right),  
\end{equation}
where in the case of spin period: $x\rightarrow P$, $\mu=1.5$, $\sigma^2=0.58$ and in the case of luminosity: $x \rightarrow L$, $\mu = -1.1$ and $\sigma^2=0.9$. 
We then assigned flux densities and DMs by sampling the corresponding distance PDFs of each target and using the \texttt{NE2001} model \citep{NE2001} respectively.
Distance estimates were drawn from  likelihood distributions with the properties given in Table~\ref{tab:object_priors}, as described in  Section~\ref{sec:target_selection}.
Finally, we adopted a beaming fraction of 70\% based on the results of \cite{1998kramer}, to decide whether a certain NS would appear as an MSP. 
The radio flux of each MSP was then compared with the flux limit of our survey at the corresponding spin period and DM (Figure\,\ref{fig:det_plots}).

This simulation allowed us to infer the posterior probability of each target to host a NS. Figure~\ref{fig:det_plots} shows the subset of samples that correspond to NSs. Those plotted in green correspond to pulsars that should have been detected by our survey and are therefore ruled out. 
To infer the posterior distribution for $f_{\rm NS}$, we run a second simulation in which we counted the number of NSs in a set comprising of one randomly selected sample per target from the above posteriors. 
Figure\,\ref{fig:pdf_8_28} shows the inferred PDF for  $f_{\rm NS}$ using $10^6$ iterations. 

\begin{figure}
    \centering
    \includegraphics[width=\columnwidth]{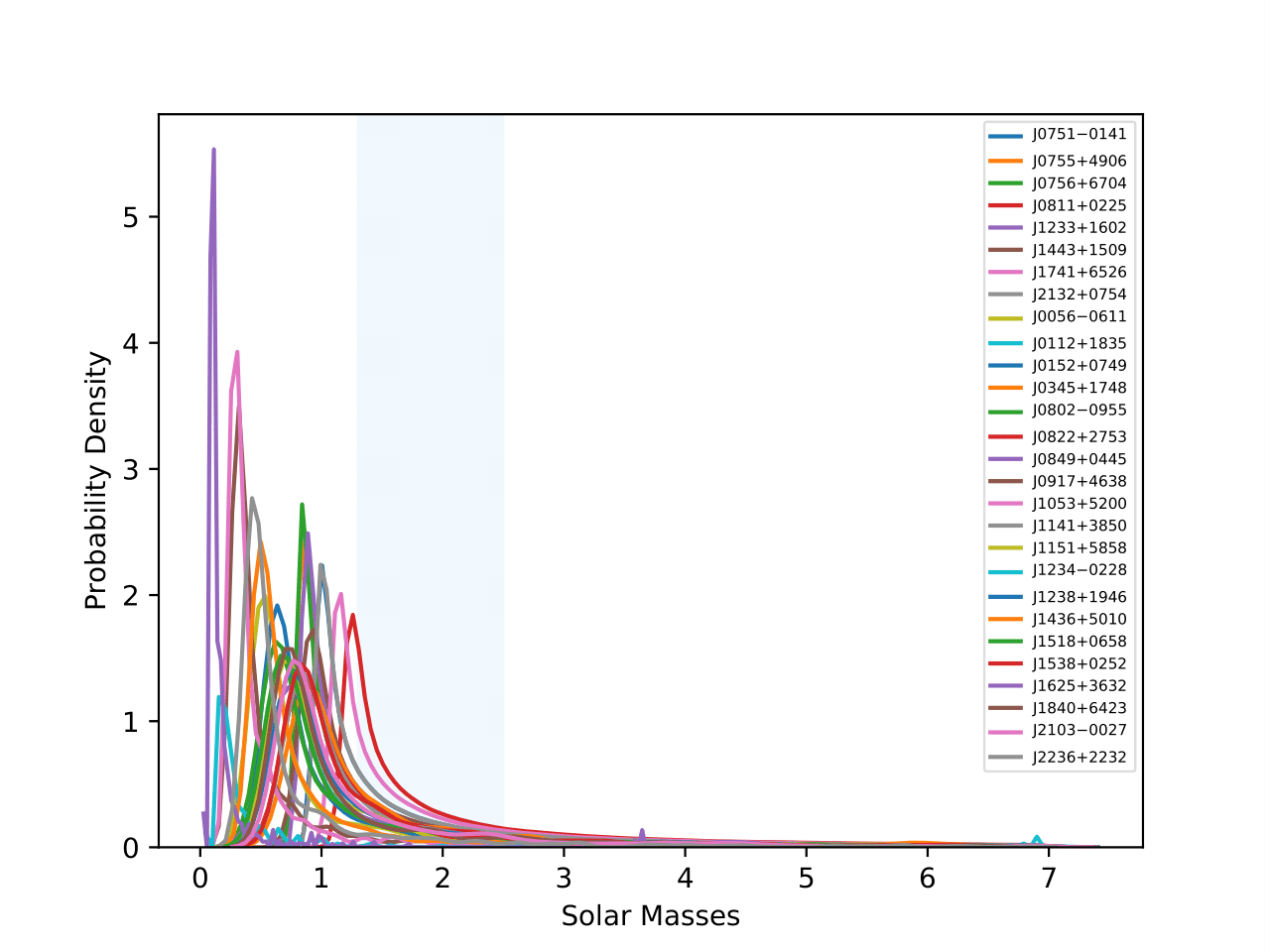}
    \caption{Prior Probability Density Functions (PDFs) for the companion mass calculated for 28 LMWD binaries observed by Effelsberg or/and GBT. The shadowed area emphasizes the mass range for which a companion is assumed to be a NS.}
    \label{fig:pdfs_28}
\end{figure}

Considering the eight LMWD systems observed with Effelsberg  (J0755+4800 was excluded as a higher-mass CO WD), we are able to constrain the fraction of LMWDs orbiting NSs in our sample to be $f_{\rm NS}\leq 0.39 $ at the 
95\% confidence limit (C.L.), where the uncertainty is dominated by Poisson noise (at the $12\%$ level), due to the small number systems in our sample. 
To improve on this estimate, we expanded our sample by considering an additional 20 LMWD binaries observed with the Green Bank Telescope \citep[GBT; ][]{2020brown}.  These observations were conducted with the GUPPI backend at a central frequency of 300\,MHz, with a detection threshold of 0.4 mJy kpc$^2$. Besides these additional sources, six of the targets considered here were also covered by this survey. 
Table~\ref{tab:object_priors} lists all targets along with the IDs of their Gaia counterparts. The m$_{\rm c}$ PDFs for these systems are shown in Figure\,\ref{fig:pdfs_28}.

To include these systems in our analysis, we followed the same procedure as above. The luminosity at 300\,MHz was calculated by scaling the \citet{Lorimer2015} relation (which is valid in the  L-band) assuming an average pulsar spectral index of -1.8 \citep{2000maron}. For the six objects that have been observed by both surveys, we took into account the most sensitive observation.

Using this extended sample, we obtain $f_{\rm NS}\leq 0.17$ at the 95\% C.L. (Figure\,\ref{fig:pdf_8_28}) using the same assumptions as above. \cite{2020brown} also presented x-ray follow up observations of their targets which further constrain the presence of a NS among the companions of these LMWDs. 
Assuming that any pulsar in that sample would have been detected regardless of distance and radio flux, yields  $f_{\rm NS}\leq 0.16$ at the 95\% C.L. (Figure \ref{fig:pdf_28_inf_sens}). 
The sensitivity of these estimates to the underlying assumptions is discussed more detail in the following section.

\section{Discussion}
\label{sec:discussion}
In this work, we present the results of a targeted  survey for pulsars orbiting  nearby WDs (on a sample of eight WDs with compact orbits and  companions with masses $\gtrsim 0.8 $\,M$_{\odot}$). 
We analyzed the data using  standard search techniques but did not detect any pulsars. Our result is complementary to the GBT radio follow-up of similar binary LMWD systems \citep{2020brown} that has also resulted in no pulsar detection.

In Section\,\ref{sec:results}, we combined the results of two surveys to constraint the fraction of LMWDs with NS companions within this particular sample to be $f_{\rm NS}\leq 0.17$ at 95\% CL. Given that both surveys targeted predominantly LMWDs with higher-mass companions, the intrinsic fraction of NSs orbiting LMWDs is likely much lower. Indeed if one considers the mass functions of all binary LMWDs found by the ELM survey \citet{2020brown}, then $f_{\rm NS}\leq 0.02$.

\begin{figure*}
\centering
\begin{tabular}{ccc}
\subfloat[]{\includegraphics{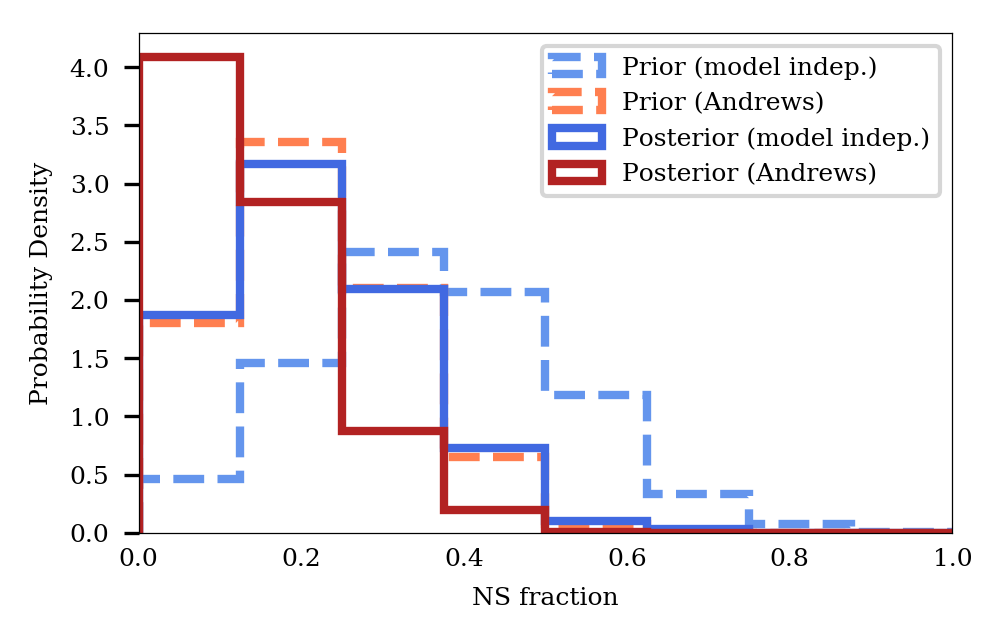}} &
\subfloat[]{\includegraphics{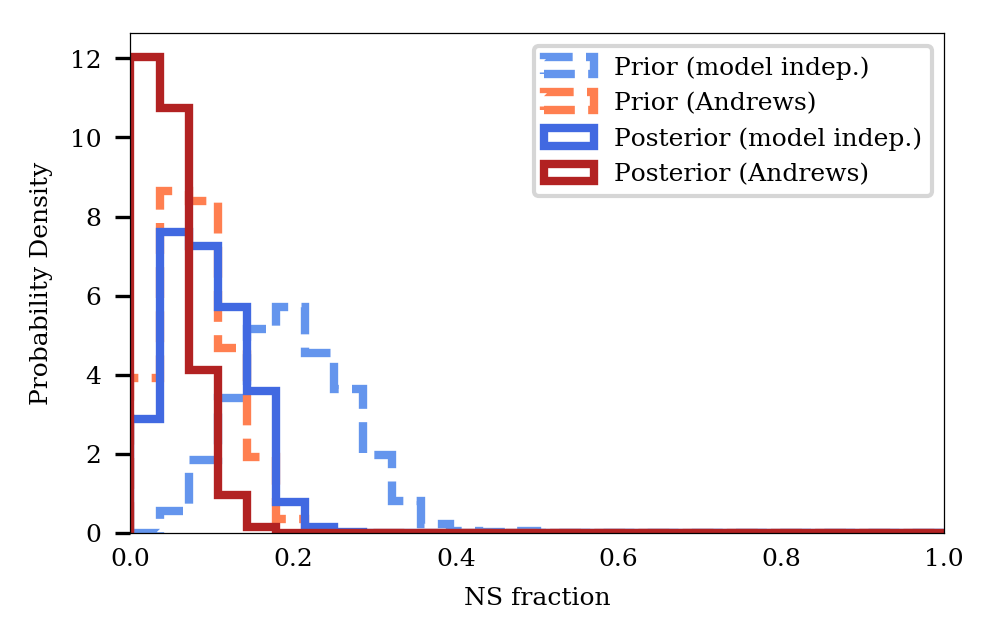}}\\
\end{tabular}
\caption{Prior and posterior Probability Density (PDF) for the NS fraction, based on (a) 8 binary LMWDs that we observed with Effelsberg radio telescope and (b) 28 binary LMWDs (8 observed with Effelsberg radio telescope and 20 by GBT). The corresponding upper limits on the neutron star fraction are $f_{\rm NS}\lesssim 0.39$ and $f_{\rm NS}\lesssim 0.17$ at 95\% CL respectively.
For comparison, for each case we used a second approach to calculate the prior probability calculated with a two Gaussian model described in \citet{Andrews2014}. We elaborate more on this model on Section\,\ref{sec:discussion}. For this model, the upper limits on the neutron star fraction are $f_{\rm NS}\lesssim 0.25$ and $f_{\rm NS}\lesssim 0.10$ at 95\% CL respectively. }
\label{fig:pdf_8_28}
\end{figure*}

Considering the possible astrophysical interpretation of this  result, it is worth 
investigating its dependence on  the underlying assumptions. 
Perhaps the most important of these is the adoption of a sharp mass cut-off between different types of compact objects. More specifically, we assumed that all stars in the 1.3\ldots2.5\,M$_{\odot}$ range are NSs, while less/more massive objects are WDs/BHs respectively. 
This assumption may not be realistic given the accumulating evidence for MSPs with masses below 1.3\,M$_{\odot}$ and  the current constraints on the nuclear EoS that place the upper NS mass limit closer to $\sim$2.1\,M$_{\odot}$ \citep{2016antoniadis,Ozel:2016oaf,Cromartie:2019kug}.  Related to this, there is likely a substantial overlap between the WD and MSP mass distributions, which could also influence our result. 
\cite{Andrews2014} employed a Gaussian mixture model to approximate the mass distribution of companions to WDs discovered in the ELM survey. 
As can be seen in Table\,\ref{tab:object_priors}, this model  predicts substantially smaller P$_{\rm prior}^{\rm NS}$ values for all sources. Implementing these priors, we infer $f_{\rm NS}\lesssim 0.10$ at 95\% CL (Figure~\ref{fig:pdf_8_28}), which is only marginally more stringent than the prior constraint.

\begin{center}
\begin{figure}
	\includegraphics[width=\columnwidth]{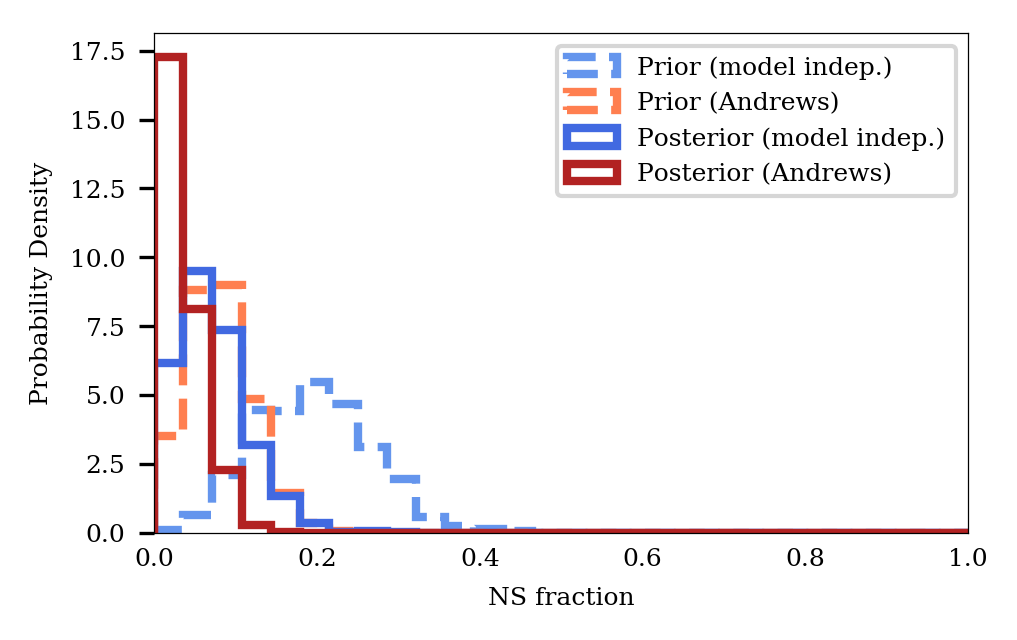}
    \caption{Prior and posterior Probability Density (PDF) for the NS fraction, assuming that in the samples of 28 LMWDs all pulsar companions would be detectable regardless of distance and radio flux and considering a beaming fraction of 70\%. For comparison a second approach to calculate the prior probability calculated with a two Gaussian model described in \citet{Andrews2014}. The upper limits on the neutron star fraction would be $f_{\rm NS}\lesssim 0.16$ and $f_{\rm NS}\lesssim 0.7$ at 95\% CL respectively.} 
    \label{fig:pdf_28_inf_sens}
\end{figure}
\end{center}

Another source of uncertainty in our estimates is the MSP luminosity distribution.
The impact of this quantity can be seen in Figure\,\ref{fig:pdf_28_inf_sens} that shows constraints on $f_{
\rm NS}$ assuming infinite sensitivity (or equivalently, extremely high intrinsic luminosity). 
At the other extreme, if MSPs are intrinsically significantly less luminous than what we assumed for our calculations, the prior and posterior distributions would be identical. 

The estimates on $f_{\rm NS}$ suggest that most compact companions to LMWDs, even those  with masses above $0.8$\,M$_{\odot}$, are CO- or ONeMg WDs. This results provides useful constraints for 
 population synthesis models \citep[e.g.][]{2014Toonen}, as it depends on a number of uncertain parameters such as the magnitude and dispersion of supernova kicks and the initial mass function and separation of the progenitor systems.

Our result is consistent with previous constraints on the fraction  of LMWD binaries with NS companions. Even though our observations suggest that this fraction is intrinsically small, targeted surveys such as the one presented here, could still provide a viable way of discovering binary MSPs. 
For instance, {\emph Gaia} has already identified more than 30,000 LMWDs \citep{2019fus}. Even a very small $f_{\rm NS}$ would imply a significant number of possible MSP discoveries. Because of the information provided by {\emph Gaia}, such systems have the potential to contribute a significant number of NS mass measurements and to improve significantly the sensitivity of Pulsar Timing Arrays \citep{Antoniadis:2020gos,2003Jaffe}.

\section*{Acknowledgements}
JA was partly supported by the Stavros Niarchos Foundation (SNF) and the Hellenic Foundation for Research and Innovation (H.F.R.I.) under the 2nd Call of ``Science and Society'' Action Always strive for excellence – ``Theodoros Papazoglou'' (Project Number: 01431). 
We are grateful to Natalya Porayko, Eleni Graikou, Tasha Gautam, Vishnu Balakrishnan, Shalini Sengupta, Savvas Chanlaridis and Pavlos Gaintatzis for discussions. 
This work relies on data from the
European Space Agency (ESA) mission {\it Gaia} (\url{https://www.cosmos.esa.int/gaia}), processed by the
{\it Gaia} Data Processing and Analysis Consortium (DPAC, \url{https://www.cosmos.esa.int/web/gaia/dpac/consortium}). 
This research made use of NumPy \citep{harris2020array}, Matplotlib \citep{Hunter:2007} and  Astropy (\url{http://www.astropy.org}), a community-developed core Python package for Astronomy \citep{Robitaille:2013mpa,Price-Whelan:2018hus} and of NASA's Astrophysics Data System Bibliographic Services.

\section*{DATA AVAILABILITY}
The data underlying this paper will be shared on a reasonable request to the corresponding author.

\onecolumn

\begin{table}
\renewcommand{\arraystretch}{1.2} 
\begin{center}
\begin{tabular}{lllllllllllll}
Name    & Gaia ID   & $P_{\rm orb}$   & $K$       & $M_{\rm wd}$  & $ M_{\rm c}$ & mass function & $P^{\rm A}_{\rm prior}$ & $P^{\rm B}_{\rm prior}$ & parallax  & distance          \\
             & & (days)    & (km/s)       & (M$_{\odot}$) & (M$_{\odot}$) & (M$_{\odot}$) &  &  & (mas) & (kpc)           \\
\hline
EFF. \& GBT \\
\hline
J0751--0141  & 3082238681040617088 & 0.08   & 432.6   & 0.17   & 0.94 & 0.671 $\pm$ 0.026   & 0.26  & 0.01    & 0.56   $\pm$ 0.10       & $1.96_{-0.33}^{+0.48}$ \\
J0755+4800$^*$   & 933556290482790528  & 0.55   & 194.5   & 0.42    & 0.9 & 0.420 $\pm$ 0.04    & 0.24  & 0.07    & 5.46  $\pm$ 0.06    & $0.183_{0.002}^{+0.002}$ \\
J0811+0225   & 3090821984198225664 & 0.82   & 220.7   & 0.17   & 1.2  & 0.916 $\pm$ 0.031   & 0.45  & 0.45    & 0.54  $\pm$ 0.20        & 
$(1.78 \pm 0.60)$ \\
J1233+1602   & 3933774668259316224 & 0.15   & 336.0   & 0.17   & 0.86 & 0.593 $\pm$ 0.021   & 0.25  & 0.06    & 1.48 $\pm$ 0.57    & $0.56_{-0.03}^{+0.05}$ \\
J1443+1509   & 1186055424948978560 & 0.19   & 307.0   & 0.17   & 0.83 & 0.570 $\pm$ 0.07    & 0.21  & 0.05    & 1.41  $\pm$ 0.17    & $0.74_{-0.09}^{+0.11}$ \\
J1741+6526   & 1633145818062780544 & 0.06   & 508.0   & 0.17   & 1.10  & 0.830 $\pm$ 0.02    & 0.37  & 0.29    & 0.86  $\pm$ 0.14    & $1.28_{-0.22}^{+0.32}$ \\
J2132+0754   & 1740741380258586624 & 0.25   & 297.3   & 0.17   & 0.95 & 0.682 $\pm$ 0.021   & 0.27  & 0.10    & 0.81  $\pm$ 0.13    & $1.34_{-0.20}^{+0.31}$ \\
\hline
ONLY EFF. \\
\hline
J0755+4906   & 933997710041265664  & 0.06   & 438.0   & 0.17   & 0.81 & 0.549 $\pm$ 0.026   & 0.23  & 0.05    & -0.44 $\pm$ 0.84   & 
$(2.47 \pm 2.00)$ \\
J0756+6704   & 1095358811015024384 & 0.62   & 204.2   & 0.18   & 0.82 & 0.545 $\pm$ 0.013   & 0.23  & 0.05    & 0.48   $\pm$ 0.05    & $2.15_{-0.19}^{+0.23}$ \\
\hline
ONLY GBT \\
\hline
J0022+0031$^*$   & 2546819845937777536 & 0.49   & 80.8    & 0.46 	 & 0.23 & 0.027 $\pm$ 0.0019  & 0.02  & 0.02    & 1.58  $\pm$ 0.34    
& $(0.51 \pm 0.04 )$ \\
J0022--1014$^*$  & 2425129334949091840 & 0.08   & 145.6   & 0.38   & 0.21 & 0.026 $\pm$ 0.0031  & 0.02  & 0.02    &   0.35 $\pm$ 0.68   & 
$(0.54 \pm 0.03)$ \\
J0056--0611  & 2524390053545665408 & 0.04   & 376.9   & 0.18   & 0.46 & 0.241 $\pm$ 0.005   & 0.10  & 0.02    & 1.60  $\pm$ 0.11    & $0.63_{-0.04}^{+0.05}$ \\
J0112+1835   & 2786627626922933248 & 0.15   & 295.3   & 0.16   & 0.62 & 0.392 $\pm$ 0.008   & 0.15  & 0.03    & 1.32  $\pm$ 0.11    & $0.78_{-0.06}^{+0.08}$ \\
J0152+0749   & 2568823856748296832 & 0.32   & 217.0   & 0.17   & 0.57 & 0.342 $\pm$ 0.009   & 0.14  & 0.02    & 1.02  $\pm$ 0.19    & $1.15_{-0.24}^{+0.37}$ \\
J0345+1748   & 44582933060466560   & 0.24   & 273.4   & 0.22   & 0.8  & 0.498 $\pm$ 0.0027  & 0.19  & 0.05    & 5.51  $\pm$ 0.06    & $0.181_{-0.002}^{+0.00}$ \\
J0802--0955  & 3038227876276093568 & 0.55   & 176.5   & 0.20   & 0.57 & 0.312 $\pm$ 0.024   & 0.13  & 0.02    & 1.00  $\pm$ 0.25    & $(1.10\pm 0.40)$ \\
J0822+2753   & 683323249480177408  & 0.24   & 271.1   & 0.19   & 0.78 & 0.500 $\pm$ 0.05    & 0.21  & 0.04    & 1.7   $\pm$ 0.17    & $0.61_{-0.06}^{+0.07}$ \\
J0849+0445   & 581901582514764928  & 0.08   & 366.9   & 0.18   & 0.65 & 0.403 $\pm$ 0.015   & 0.15  & 0.03    & 0.56   $\pm$ 0.29   & $(1.55\pm 0.70)$ \\
J0917+4638   & 1011260873161111040 & 0.32   & 148.8   & 0.17   & 0.28 & 0.108 $\pm$ 0.004   & 0.05  & 0.02    & 0.45  $\pm$ 0.23    & $(2.23\pm 0.50)$ \\
J1005+0542$^*$   & 3873389833259088384 & 0.31   & 208.9   & 0.38 	 & 0.7  & 0.289 $\pm$ 0.028   & 0.14  & 0.03    & 0.60 $\pm$ 0.71  & $(0.61\pm 0.08)$\\
J1053+5200   & 837162724550333312  & 0.04   & 264.0   & 0.20   & 0.26 & 0.081 $\pm$ 0.0018  & 0.04  & 0.02    & 1.51  $\pm$ 0.44    & $(1.76\pm 0.40)$\\
J1056+6536$^*$   & 1058999022339181440 & 0.04   & 267.5 	& 0.33   & 0.34 & 0.086 $\pm$ 0.007   & 0.05  & 0.02    & 0.66  $\pm$ 0.38 & $(0.70 \pm 0.10)$\\
J1104+0918$^*$   & 3866880552624195584 & 0.55   & 142.1 	& 0.46 	 & 0.55 & 0.165 $\pm$ 0.021   & 0.13  & 0.02    & 5.30  $\pm$ 0.08    & $0.188_{-0.002}^{+0.003}$ \\
J1141+3850   & 766520476855004672  & 0.26   & 265.8   & 0.18   & 0.77 & 0.505 $\pm$ 0.020   & 0.21  & 0.04    & 0.66  $\pm$ 0.25    & $(1.66 \pm 0.45)$\\
J1151+5858   & 846536370415062784  & 0.67   & 175.7   & 0.19   & 0.63 & 0.380 $\pm$ 0.04    & 0.15  & 0.03    & 1.07  $\pm$ 0.42    &  $(0.53\pm 0.05)$\\
J1234--0228   & 3683189405578704640 & 0.09   & 94.0    & 0.23   & 0.09 & 0.008 $\pm$ 0.0007  & 0.01  & 0.02    & 1.28  $\pm$ 0.14    & $0.81_{-0.08}^{+0.12}$ \\
J1238+1946   & 3948319763985443200 & 0.22   & 258.6   & 0.21   & 0.68 & 0.399 $\pm$ 0.012   & 0.18  & 0.03    & 0.45  $\pm$ 0.10    & $(1.96\pm 0.40)$\\
J1436+5010   & 1603554764703627520 & 0.04   & 347.4   & 0.23   & 0.45 & 0.199 $\pm$ 0.015   & 0.09  & 0.02    & 1.05  $\pm$ 0.13    & $1.01_{-0.12}^{+0.16}$ \\
J1518+0658   & 1163266672074075008 & 0.60   & 172.0   & 0.22   & 0.6  & 0.321 $\pm$ 0.011   & 0.15  & 0.02    & 2.86  $\pm$ 0.09    & $0.35_{-0.01}^{+0.01}$ \\
J1538+0252   & 4424162321742140928 & 0.42   & 227.6   & 0.17   & 0.76 & 0.512 $\pm$ 0.033   & 0.20  & 0.04    & 0.71  $\pm$ 0.20    & $(1.58\pm 0.50)$\\
J1625+3632   & 1329610343131728896 & 0.23   & 58.4    & 0.2    & 0.07 & 0.005 $\pm$ 0.0011  & 0.01  & 0.02    & 0.40  $\pm$ 0.23    & $(2.22\pm 0.90)$\\
J1630+4233$^*$   & 1405204172723046144 & 0.03   & 295.5   & 0.30   & 0.3  & 0.074 $\pm$ 0.004   & 0.05  & 0.02    & 1.17 $\pm$ 0.19 & $0.94_{-0.16}^{+0.24}$ \\
J1840+6423   & 2256447694151327488 & 0.19   & 272.0   & 0.18   & 0.65 & 0.399 $\pm$ 0.009   & 0.18  & 0.03    & 1.30 $\pm$ 0.16     & $0.81_{-0.10}^{+0.13}$ \\
J2103--0027   & 2690059032483402368 & 0.20   & 281.0   & 0.16   & 0.7  & 0.467 $\pm$ 0.016   & 0.17  & 0.03    &  0.93 $\pm$ 0.25   & $(1.21\pm 0.40)$\\
J2236+2232   & 1874523804732334464 & 1.01   & 119.9   & 0.19   & 0.39 & 0.180 $\pm$ 0.009   & 0.10  & 0.02    & 2.73  $\pm$ 0.08    & $0.37_{-0.01}^{+0.01}$ \\
\hline
\end{tabular}
\end{center}
\caption{The objects observed by Effelsberg Radio Telescope (EFF) and/or Green Bank Telescope (GBT). Marked objects (*) were excluded from the NS fraction simulations due to high mass of the white dwarf. The orbital period ($P_{\rm orb}$), radial velocity ($K$) and the mass of the LMWDs ($M_{\rm wd}$) are provided by \citet{brown2013, 2020brown}. Minimum companion mass ($M_{\rm c}$) and the mass function ($f$) was calculated as we described in Section \ref{sec:target_selection}. $P^{\rm A}_{\rm prior}$ and $P^{\rm B}_{\rm prior}$ are different calculations of the prior probability for each system to host a NS based on its mass function. The first one is based on the model independent calculation that we describe in Section \ref{sec:target_selection} and the second on the work of \citet{Andrews2014}. Parallaxes are provided by Gaia EDR3. Most distances, provided with two-sigma errors, were calculated as we describe in Section \ref{sec:target_selection}. For non significant or non-valid parallax values the photometric distances were used. Photometric distances are listed inside parentheses and are provided with the symmetric one-sigma error bounds that was used in the calculation.}
\label{tab:object_priors}
\end{table}


\newpage
\twocolumn
\bibliographystyle{mnras}
\bibliography{mnras} 




\bsp	
\label{lastpage}
\end{document}